\documentclass[11pt, a4paper]{article}
\usepackage{amssymb,moriond,epsfig}
\usepackage[polish,english]{babel}

\def\be{\begin{equation}}
\def\ee{\end{equation}}
\def\bea{\begin{eqnarray}}
\def\eea{\end{eqnarray}}

\def\dzero{D\O}
\newenvironment{2figures}[1]{\begin{figure}[#1]
  \begin{center}
    \begin{tabular}{p{.48\textwidth}p{.48\textwidth}} }
 {  \end{tabular}
  \end{center}\vspace*{-8mm}
 \end{figure}
}

\begin{document}
\vspace*{4cm}
\title{LATEST JET RESULTS FROM THE TEVATRON}
\author{ M. {\selectlanguage{polish} "CWIOK}}
\address{%
 (on behalf of CDF and \dzero~collaborations)\linebreak %
 University College Dublin, %
 School of Physics, %
 Belfield, Dublin 4, Ireland %
} %
%
\maketitle\abstracts{%
Recent QCD jet production measurements in $p\bar{p}$ collisions at
$\sqrt{s}=1.96\,$TeV at the Tevatron
Collider at Fermilab are presented. %
Preliminary: %
inclusive jet, dijet, isolated photon~+~jet and $Z$~+~jets %
measurements are compared to available perturbative QCD models. %
}%
%
The production of particle jets with high transverse momenta in
hadronic collisions is described in perturbative Quantum
Chromodynamics (pQCD) as resulting from the hard scattering of
strongly interacting constituents of the colliding hadrons.

%
%
%
Inclusive jet rates observed in hadronic collisions at high values
of transverse momenta provide a basic test of pQCD. %
The \dzero~and CDF collaborations\,\cite{d0:detector,cdf:detector}
have measured the inclusive jet production cross section using {\em
midpoint cone} and $k_T$ algorithms\,\cite{kt_cone} using data
corresponding
to the integrated luminosities of about $1\,{\rm fb}^{-1}$. %
The \dzero~result\,\cite{d0note_incljet} is shown in Figure~\ref{fig:d0_inclusive} %
for two regions of rapidity\,%
\footnote{%
The rapidity $y$ is defined as $y=-{1\over
2}\ln{{E+p_z}\over{E-p_z}}$ where $E$ and $p_z$ denote the energy
and the momentum component along the proton beam direction, respectively.} %
(closed and open circles).
The error bars correspond to the total measurement uncertainty. %
The data are corrected for the jet energy scale (JES) determined
from isolated photon plus jets events, selection efficiencies and
migrations due to $p_T$ resolution (an ansatz function convoluted
with the jet $p_T$ resolution measured directly in data). %
The JES is the dominant source of systematic uncertainty. %
The integrated luminosity is known with accuracy of 6\%. %
The data are compared to the next-to-leading order (NLO) pQCD
predictions computed using NLOJET++\,\cite{NLOJET} with parton
density functions (PDFs) from CTEQ6.1M,\cite{cteq6M} after applying
threshold corrections at 2-loop (next-to-next-leading-logarithm)
accuracy.\cite{kidonakis} The same jet algorithm was used in the
calculations and the pQCD predictions are also corrected for
hadronization effects using PYTHIA.\cite{PYTHIA}
%
%
\begin{2figures}{t}
\resizebox{0.97\linewidth}{!}{
\includegraphics*[0,5][567,527]{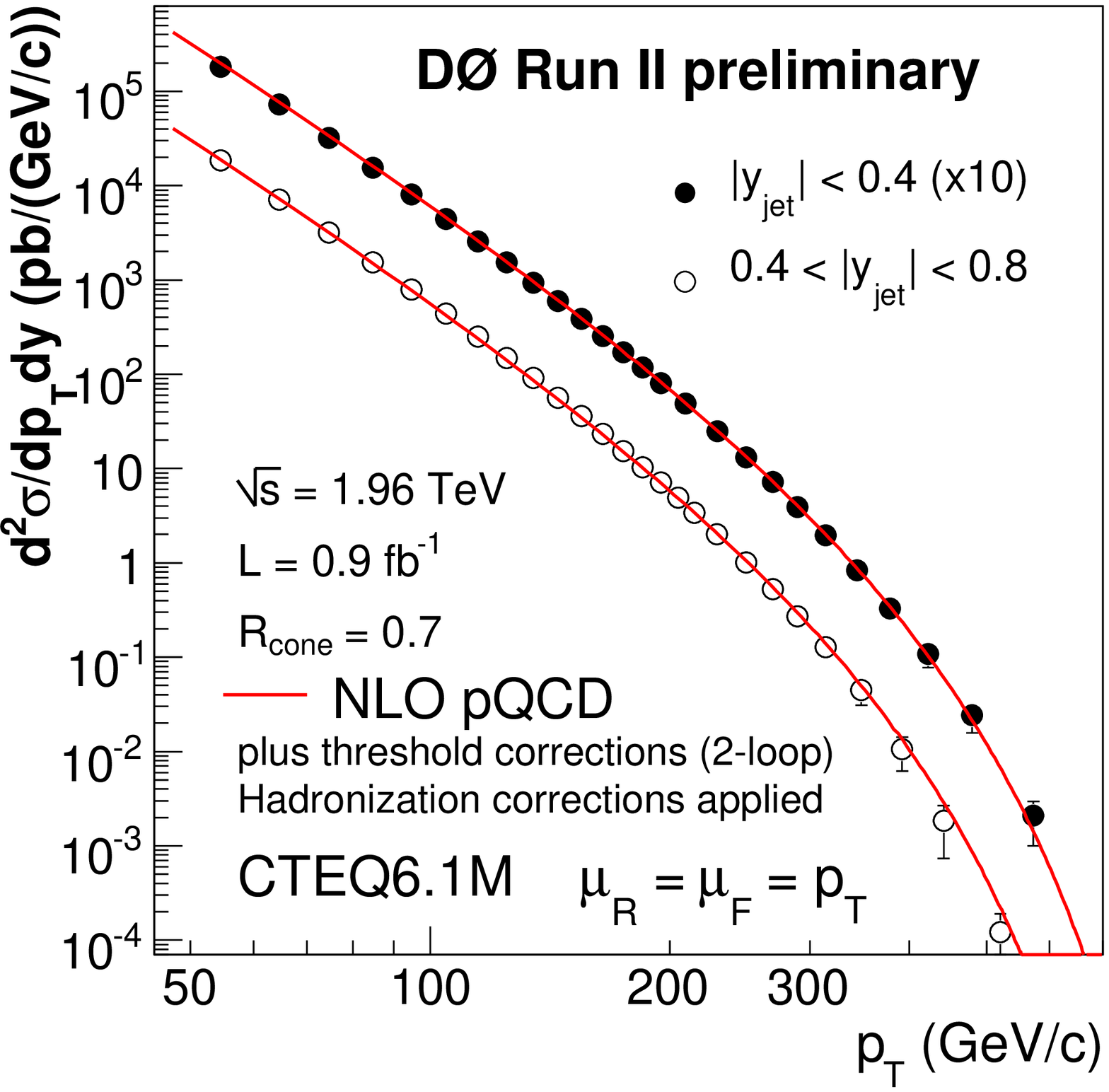}} & %
\resizebox{0.97\linewidth}{!}{
\includegraphics*[0,-10][550,500]{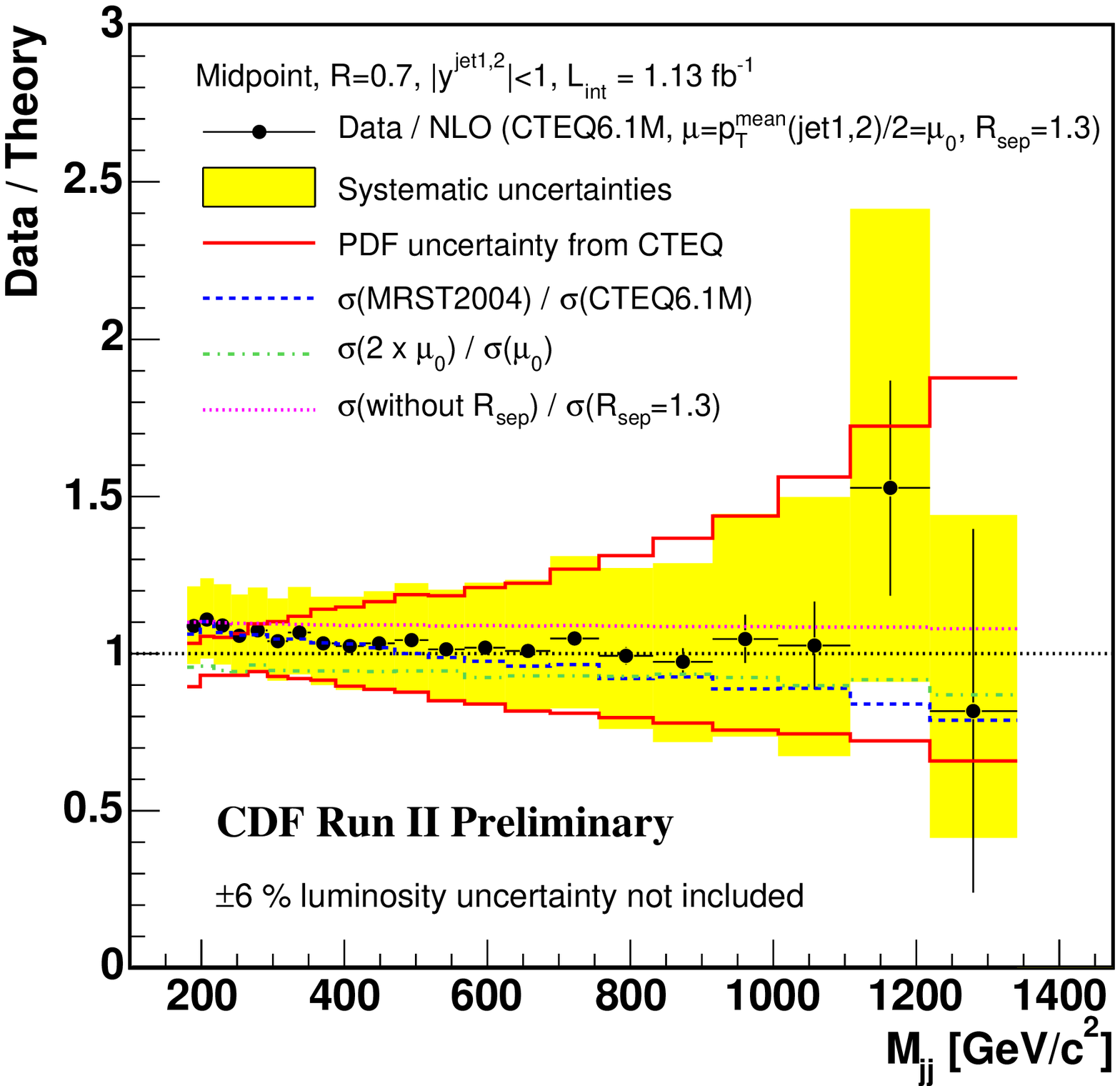}}\\[-5mm]
\caption{The inclusive jet differential cross section measured in
two regions of jet rapidity.}
\label{fig:d0_inclusive} & %
\caption{The data/theory ratio for dijet inclusive cross sections as
a function of the dijet invariant mass.} \label{fig:cdf_dijet}
\end{2figures}
%
%
%
The theory describes the data well over the whole measured $p_T$
range in all rapidity regions. The experimental uncertainties are
competitive with those from the proton PDFs and the data therefore
further constrain the gluon density functions at high-$x$.
%
Inclusive jet spectra measured by CDF are also in good agreement
with the NLO pQCD predictions.\cite{Norniella}

%
%
%
The rate of dijet event produced in hadronic collisions not only
provides a test of pQCD but also is sensitive to new physics such as
compositeness and massive particles decays. %
The ratio of the dijet cross section measured by CDF to theory is
shown as a function of dijet invariant mass ($M_{jj}$) in
Figure~\ref{fig:cdf_dijet}. %
The measurement corresponds to 1.13$\,$fb$^{-1}$ and centrally produced jets. %
The jets were selected using midpoint cone algorithm.\cite{kt_cone} %
The error bars and shaded bands represent the statistical and
systematic uncertainties respectively. %
Theoretical predictions were calculated using
NLOJET++\,\cite{NLOJET} with PDFs from CTEQ6.1M\,\cite{cteq6M} and
corrected to the hadron level. %
The systematic errors are comparable to the PDF uncertainties and
NLO pQCD predictions are consistent with the data over the whole
measured $M_{jj}$ range.

%
%
%
The production rates of $b\bar{b}$ jet pairs have also been studied
by CDF using a data sample of $260\,{\rm pb}^{-1}$. Such a
measurement provides insight into $b$-quark direct production,
flavour excitation and gluon splitting mechanisms and also allows a
test of radiative gluon corrections.
The selected events were first required to contain two jets with
transverse energy above 20\,GeV associated to two displaced vertex
tracks at the trigger level. A Run~I cone algorithm\,\cite{kt_cone}
was used to identify the jets. Jets having a positively displaced
secondary with respect to the jet axis were tagged as {\em
``SVT~b-jets''}. Two positively tagged jets with central
pseudorapidities\,%
\footnote{Pseudorapidity $\eta$ is defined as %
          $\eta = -\ln{ \tan{ \frac{\theta}{2}}}$, %
          where $\theta$ is the polar angle w.r.t. the proton beam direction.} %
%
%
and transverse energies of a leading and a second jet above 35
and 32\,GeV respectively were required. %
The invariant mass of the tracks associated to the secondary vertex
was fitted using signal and background Monte Carlo templates to
determine the $b\bar{b}$ purity of the final event sample. %
The resulting purity was about 80\%.
%
%
%
\begin{figure}[t!]\vspace{-1mm} %
\centerline{%
\resizebox{!}{0.395\linewidth}{
\includegraphics*[0,0][567,370]{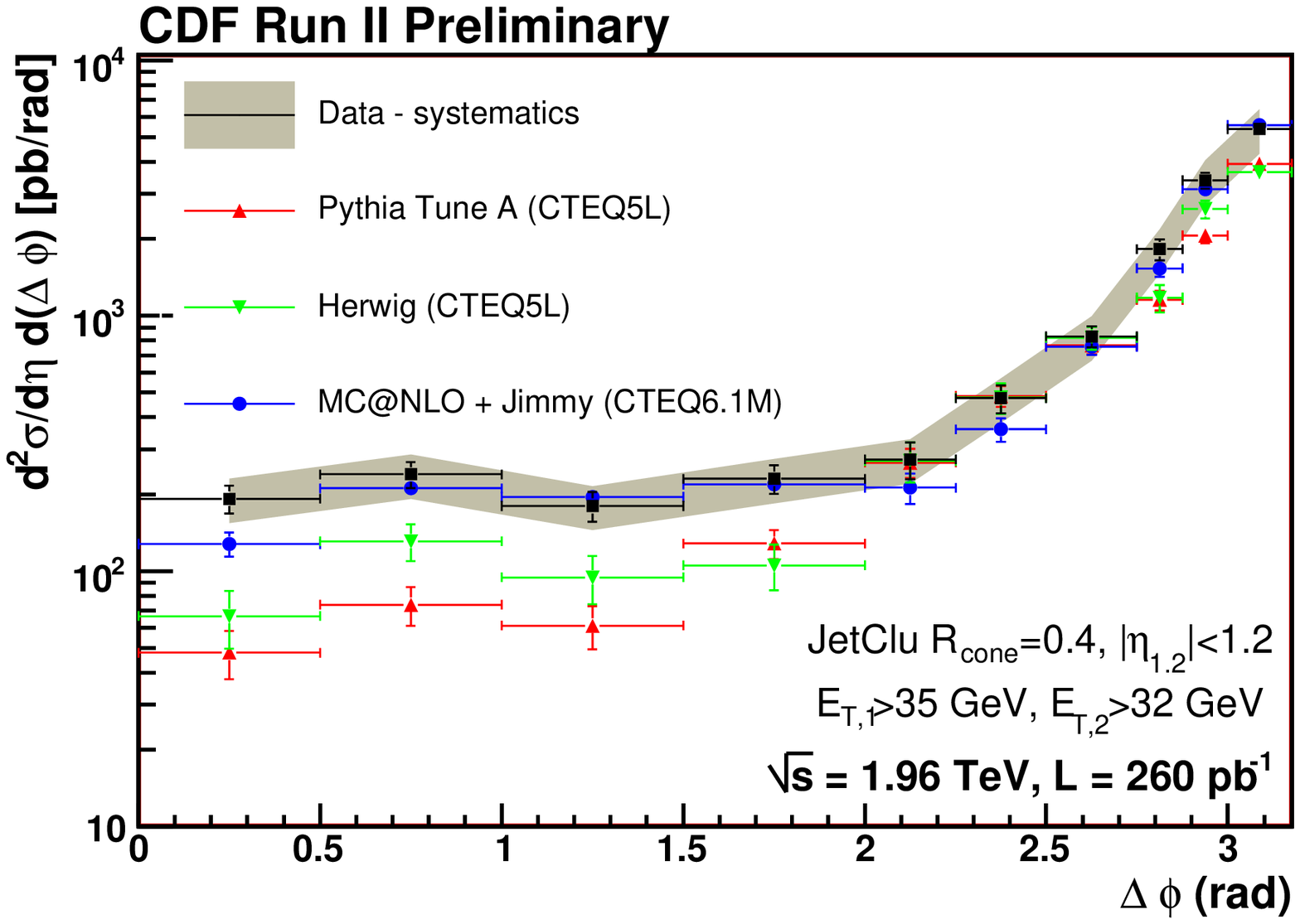}} %
}%
\vspace{-2.5mm}\caption{The differential $b\bar{b}$ cross section as
a function of the azimuthal angle between two jets.}
\label{fig:cdf_bbbar}\vspace{-4mm}
\end{figure}
%
%
The differential cross section measured as a function of the
azimuthal angle between two jets ($\Delta\phi_{jj}$), unfolded to
the hadron level, is shown in Figure~\ref{fig:cdf_bbbar} (full
squares). The error bars and shaded bands correspond to the measured
statistical and systematic uncertainties respectively.
The data are compared to three theoretical models: two predictions
at leading order (LO) from PYTHIA\,\cite{PYTHIA} (Tune~A) and
HERWIG\,\cite{HERWIG} with PDFs from CTEQ5L\,\cite{cteq5L} and a NLO
prediction from MC@NLO\,\cite{MCNLO} using CTEQ6.1M\,\cite{cteq6M}
PDFs and with multiple parton interactions simulated by
JIMMY.\cite{JIMMY}
The $\Delta\phi_{jj}$ spectrum is sensitive to contributions arising
from soft gluon radiation and therefore only the NLO pQCD model
describes data reasonably well even at large departures from a
back-to-back jet topology.

%
%
%
Photons produced directly in parton-parton QCD interactions arrive
unaltered at the electromagnetic calorimeter and carry clean
information of the dynamics of the hard scatter. %
At the Tevatron prompt photon production is dominated by the Compton
scattering subprocess \mbox{$q g \rightarrow \gamma q$} for photon
transverse momenta $p_T^{\gamma}\lesssim 150\,$GeV/c. %
The differential cross section %
$d^3\sigma/(dp_T^{\gamma}\,d\eta{^\gamma}\, d\eta^{jet})$ %
for the production of a photon and a jet measured by \dzero~using a
$1.1\,{\rm fb^{-1}}$ data sample is shown in
Figure~\ref{fig:d0_gammajet}.\cite{d0note_gammajet} %
The jets were reconstructed using a midpoint cone
algorithm\,\cite{kt_cone} and were required to have: transverse
momenta $p_T^{jet}>15\,$GeV/c and pseudorapidities either in the
central calorimeter ({\em ``CC''}, $|\eta^{jet}|<0.8$) %
or in the end cap (forward) calorimeter region %
({\em ``EC''}, $1.5<|\eta^{jet}|<2.5$). %
Photons were selected with transverse momenta
$30<p_T^{\gamma}<300\,$GeV/c and central pseudorapidities
$|\eta^{\gamma}|<1$.
Strong isolation criteria were imposed on photon candidates in order
to filter background events with neutral hadrons decaying to photons
in the final state. The purity of the resulting sample was estimated
with the help of an artificial neural network trained to distinguish
between direct photons and background.
The measured cross section is corrected for the finite resolution of
the calorimeter. Events with a leading jet and a photon contained in
the same hemisphere in terms of their pseudorapidities are denoted
as {\em ``SS''} (same sign) while the remaining ones as {\em ``OS''}
(opposite sign).
The four curves overlaid on the data represent the NLO pQCD
predictions from JETPHOX\,\cite{JETPHOX} with the choice of
CTEQ6.1M\,\cite{cteq6M} PDFs and fragmentation
functions.\cite{BFG_FF} %
The theory qualitatively reproduces the data in some kinematic
regions.
%
%
%
\begin{2figures}{t}
\resizebox{0.97\linewidth}{!}{ %
\includegraphics[17,12][536,530]{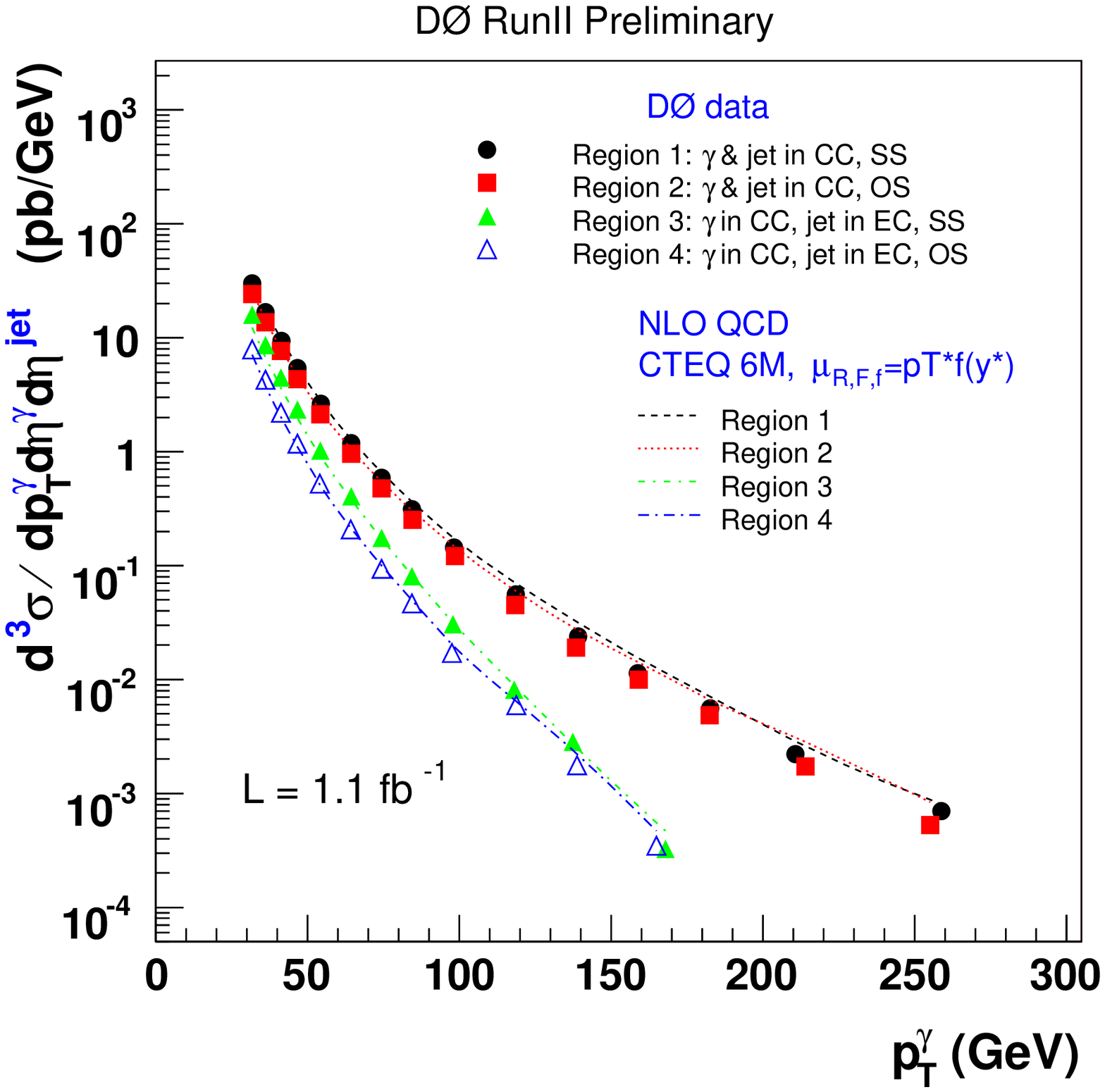} %
} & %
\resizebox{0.97\linewidth}{0.91\linewidth}{ %
\includegraphics[0,325][528,707]{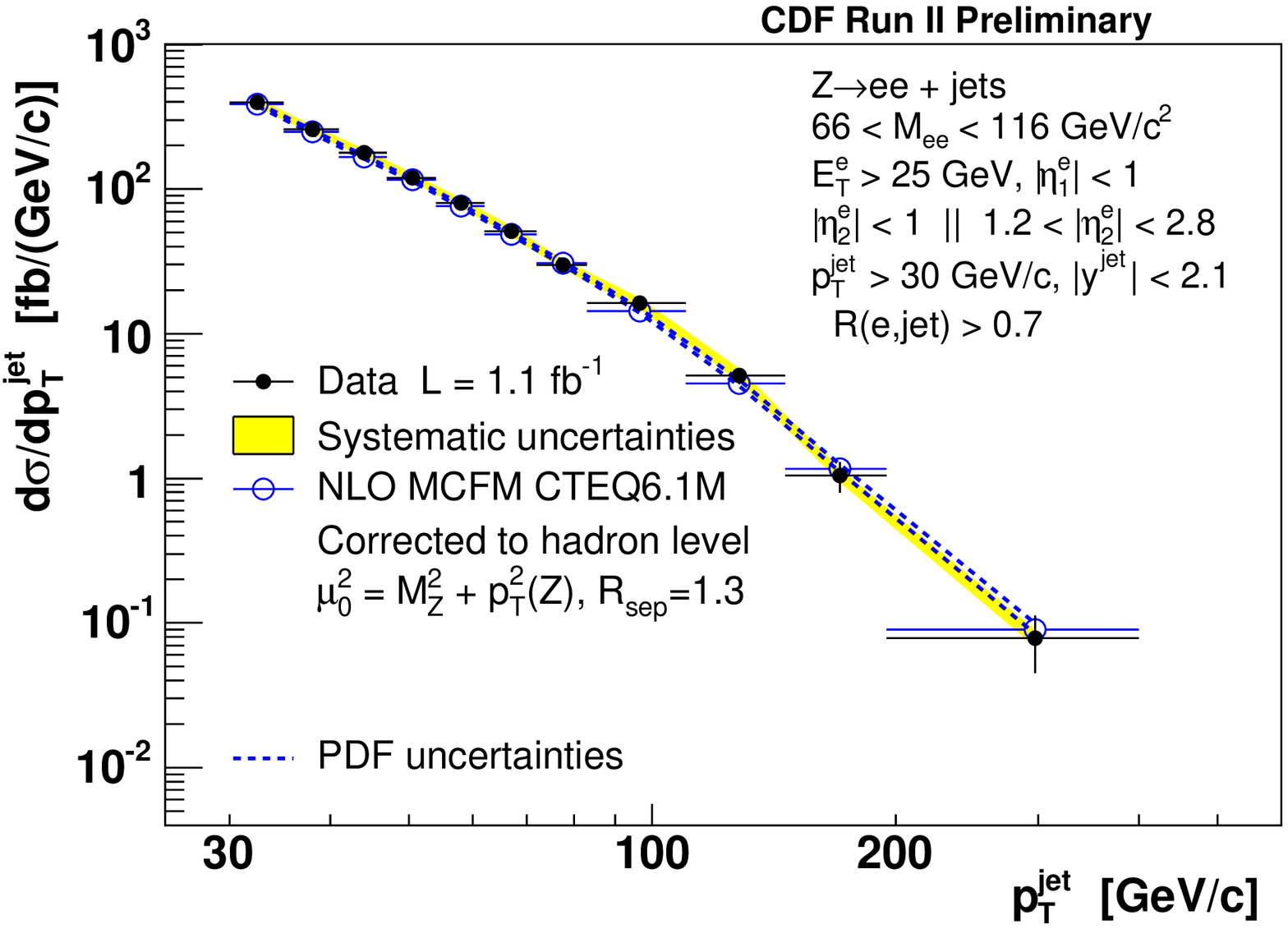} %
} \\[-5mm] %
\caption{The differential $\gamma$+jet cross sections as a function
of photon $p_T$ for four rapidity regions.} %
 \label{fig:d0_gammajet} & %
\caption{The differential $Z(\rightarrow e^{+}e^{-})\,+\,$jets cross
section as a function of jet $p_T$.} %
 \label{fig:cdf_Zjets}
\end{2figures}

%
%
%
Jets accompanied by $W$ or $Z$ vector bosons in $p\bar{p}$
collisions constitute an important background for top quark
production, Higgs and SUSY searches. %
Their production rates are also sensitive to physics beyond the
Standard Model (compositeness and decays of heavy objects). %
In addition, $Z+$jets events are suitable for testing
phenomenological models of the underlying event in $p\bar{p}$
collisions by studying integrated and differential jet shapes or
energy flow with respect to the momentum of a $Z$ boson.
The CDF collaboration studied production of $Z+$jets events with $Z$
bosons decaying into $e^{+}e^{-}$ pair using $1.1\,{\rm fb}^{-1}$ of
data. Such a channel provides much cleaner experimental signature
than $W+$jets one, albeit has 10 times smaller cross section. The
analysis covered the following kinematic region: jets reconstructed
using midpoint cone algorithm\,\cite{kt_cone} having
$p_T^{jet}>30\,$GeV/c and $|\eta^{jet}|<2.1$, electrons with
$E_T^{e\,1,2}>25\,$GeV, $|\eta^{e\,1}|<1$, $|\eta^{e\,2}|<2.8$ and
isolated from jet cones.
%
%
The acceptance window for the invariant mass of an $e^{+}e^{-}$ pair
was taken to be 66 to $116\,{\rm GeV/c^2}$ to suppress background.
The cross section as a function of the transverse momentum of a
leading jet is shown in Figure~\ref{fig:cdf_Zjets} (closed circles)
and is compared to the NLO prediction using MCFM\,\cite{MCFM} with
CTEQ6.1M\,\cite{cteq6M} PDFs after corrections to the hadron level
(open circles). The data/theory ratio is consistent with unity,
although statistical errors dominate at $p_T^{jet}>100\,$GeV/c
region.

%
%
%
Present experimental data on QCD jets from the Tevatron Collider are
reasonably well described by existing next-to-leading calculations
after applying parton-to-hadron level corrections.
The CDF and \dzero~collaborations have now collected about
$2.2\,{\rm fb}^{-1}$ of data on tape and anticipate up to $8\,{\rm
fb}^{-1}$ by end of 2009. This should make possible even higher
precision tests of pQCD theory over extended kinematic regions.
%
%
%
\section*{Acknowledgments}\vspace{-1mm} %
I would like
thank the staffs at Fermilab and collaborating institutions and
acknowledge support from SFI (Ireland) and EU Marie Curie Programme.
%
%
%

\end{document}